\documentclass{article}
\usepackage{spconf,amsmath,graphicx}

\usepackage{color}
\usepackage{stfloats}
\usepackage{multirow}

\makeatletter 
  \newcommand\figcaption{\def\@captype{figure}\caption} 
  \newcommand\tabcaption{\def\@captype{table}\caption} 
\makeatother

\title{LSR: A Light-Weight Super-Resolution Method}

\name{Wei Wang \textsuperscript{1},
Xuejing Lei \textsuperscript{1},
Yueru Chen  \textsuperscript{2},
Ming-Sui Lee \textsuperscript{3},
C.-C. Jay Kuo \textsuperscript{1}
\thanks{The authors acknowledge the gift support from MediaTek Inc.
as well as the Center for Advanced Research Computing (CARC) at the
University of Southern California for providing computing resources that
have contributed to the research results reported within this
publication. URL: https://carc.usc.edu.}}
\address{University of Southern California, Los Angeles, California, USA$^1$\\
Peng Cheng Laboratory, Shenzhen, Guangdong, China$^2$\\
National Taiwan University, Taipei, Taiwan$^3$}

\begin{document}
\ninept

\maketitle

\begin{abstract}

A light-weight super-resolution (LSR) method from a single image
targeting mobile applications is proposed in this work. LSR predicts the
residual image between the interpolated low-resolution (ILR) and
high-resolution (HR) images using a self-supervised framework. To lower
the computational complexity, LSR does not adopt the end-to-end
optimization deep networks. It consists of three modules: 1) generation
of a pool of rich and diversified representations in the neighborhood of
a target pixel via unsupervised learning, 2) selecting a subset from the
representation pool that is most relevant to the underlying
super-resolution task automatically via supervised learning, 3)
predicting the residual of the target pixel via regression. LSR has 
low computational complexity and reasonable model size so that it can be
implemented on mobile/edge platforms conveniently. Besides, it offers
better visual quality than classical exemplar-based methods in terms of
PSNR/SSIM measures. 

\end{abstract}


\section{INTRODUCTION}\label{sec:intro}

Single image super-resolution (SISR) \cite{irani1991improving} is an
intensively studied topic in image processing. It aims at recovering a
high-resolution (HR) image from its low-resolution (LR) counterpart. 
SISR finds wide real-world applications such as remote sensing, medical
imaging, and biometric identification. Besides, it attracts attention due to
its connection with other tasks (e.g., image registration, compression,
and synthesis). 

SISR is an ill-posed problem since multiple HR patches can map to the
same LR patch. To solve this one-to-many mapping problem, SISR is
typically formulated as a regularized optimization problem or a
generative problem with supervised learning. For the former, one may
impose priors to regularize the ill-posed problem, yet the performance
improvement is limited. For the latter, there are two main approaches:
exemplar-based (or dictionary-based) methods and deep-learning (DL)
methods. 

DL-based super-resolution methods have been dominating in the field
since 2015 \cite{dong2015image}. They have been intensively studied in
the last eight years. They offer better HR images in terms of PSNR/SSIM
quality metrics at the cost of higher network parameters and larger
computational complexity. One of the main applications of the SR
techniques is mobile platforms and consumer electronics (e.g., smart
TVs). DL-based SR solutions cannot be easily implemented on
resource-constrained computational platforms due to the price
consideration. 

To address this problem, a light-weight super-resolution (LSR) method is
proposed in this work. LSR predicts the residual image between the
interpolated low-resolution (ILR) and HR images using a self-supervised
learning paradigm. LSR does not adopt the end-to-end optimization deep
networks. Instead, it consists of three cascaded modules. First, it
creates a pool of rich and diversified representations in the
neighborhood of a target pixel via unsupervised learning. Second, it
selects a subset from the representation pool that is most relevant to
the underlying super-resolution task automatically via supervised
learning. Third, it predicts the residual of the target pixel based on
the selected features through regression via classical machine learning
such as the XGBoost regressor. LSR offers visual quality that is better
than exemplar-based methods and comparable with the entry-level DL-based
SR solution, SRCNN, in terms of PSNR/SSIM measures. 

It is worthwhile to highlight the value of this work.  Our main
contributions lie in the low computational complexity of
the proposed LSR method. As presented in the experimental
section, there are two versions of the LSR method, namely, LSR V1 and
LSR V2.  Both have around 380K parameters. We use the number of
floating-point operations per pixel (FLOPs/pixel) to measure the
complexity in inference. LSR V1 and LSR V2 demand 9.28K and 3.83K FLOPs/pixel,
respectively. For complexity benchmarking, we choose two well-known and
representative DL-based SR solutions; i.e., SRCNN and VDSR.  SRCNN and
VDSR demand 114K and 1.33M FLOPs/pixel, respectively. The LSR method has
a clear advantage over DL-based solutions when being deployed on the
mobile/edge platforms.  

The rest of the paper is organized as follows.  Related work is reviewed
in Sec. \ref{sec:review}. The proposed LSR method is presented in Sec.
\ref{sec:method}. Experimental results are shown in Sec. \ref{sec:exp}.
Finally, concluding remarks and future research directions are discussed
in Sec. \ref{sec:conc}. 

\section{REVIEW OF RELATED WORK} \label{sec:review}

{\bf Exemplar-based Methods.} Image patches are viewed as examples of
local regions. Patches are partitioned and represented in the form of
dictionary atoms. Finally, proper mappings from LR-to-HR patches are
developed inside each partition. Examples include \cite{chang2004super,
yang2012coupled, timofte2014a+, huang2015single, schulter2015fast}.
Sometimes, priors are leveraged for patch partitioning
\cite{sandeep2016single}. Since the dictionary size can be expanded
flexibly, they are non-parametric methods. There are limitations with
exemplar-based methods. First, the LR-to-HR mapping is based on
hand-crafted features. There is no clear guideline in patch sizes for
mapping learning. Second, the training of the LR-to-HR mapping is
time-consuming \cite{yang2012coupled, timofte2014a+}. Last, the quality
of their enhanced SR images is inferior to that achieved by modern
DL-based methods.

{\bf DL-based Methods.} The application of DL to the SISR problem can be
traced back to SRCNN in 2015 \cite{dong2015image}. Substantial advances
have been made along this direction in the last eight years, e.g.,
\cite{kim2016accurate, zhang2018residual, dai2019second, mei2020image,
chen2021pre, kong2021classsr, wang2021exploring, zhou2019kernel,
hu2019meta}. In earlier years, the focus was on achieving higher
performance (namely, better PSNR and SSIM) \cite{dong2015image,
kim2016accurate, zhang2018residual, dai2019second, mei2020image,
chen2021pre}. Research on efficiency has been considered in recent
years, e.g., \cite{kong2021classsr, wang2021exploring}. Other
SISR-related problems have also been explored, e.g. unknown degradation
kernels \cite{zhou2019kernel}, magnification with a non-integer factor
\cite{hu2019meta}, etc. Although DL-based methods offer significant
performance breakthrough, it is a major challenge to apply them to
practical SR problems in the mobile/edge devices due to their heavy
computational and memory costs. Besides, they lack mathematical
transparency. 

{\bf Green Learning.} Green learning \cite{kuo2022green} is an emerging
learning paradigm emphasizing lower computational complexities and smaller 
model sizes. It has a modular design that consists of
three cascaded modules: 1) unsupervised representation learning, 2)
supervised feature learning, and 3) supervised decision learning. For
unsupervised representation learning, Kuo {\em et al.} interpreted the
convolution operations in convolutional neural networks (CNNs) as joint
spatial-spectral signal transforms in \cite{kuo2016understanding,
kuo2018data, kuo2019interpretable} and proposed two one-stage
data-driven transforms, the Saak transform \cite{kuo2018data} and the
Saab transform \cite{kuo2019interpretable}. To achieve multi-stage
signal transforms, Kuo developed a successive-subspace-learning (SSL)
strategy in \cite{kuo2018data, kuo2019interpretable}. For supervised
feature learning, the problem of selecting the most discriminant (or
relevant) features from the pool of rich representations for some
classification (or regression) tasks based on user's labels is examined
in \cite{yang2022supervised}. Green learning has been successfully
applied to many applications. Our LSR method follows the same pipeline
as elaborated below. 

\section{Proposed LSR METHOD} \label{sec:method}

For self-supervised SR, LR images are obtained from HR images via
bicubic down-sampling in training and test image sets. Following the
standard pipeline, we only focus on the luminance (or Y) component. As
a pre-processing step, a Lanczos interpolation is applied to LR images
to yield interpolated LR (ILR) images whose resolution is the same as HR
images. To regularize the ill-posedness of the problem, LSR uses the
neighborhood of a target pixel to predict its residue, which is the
difference between HR and ILR images at the pixel. Thus, the input and
output to the proposed LSR system are an ILR patch (of size $15\times
15$) and the residual value of its center pixel, respectively. 

It is desired to divide pixels into easy and hard two classes. Pixels
in smooth regions are easy samples. Their residual values are small
since the interpolation can predict their values quite well.
Furthermore, their residuals can be predicted using a simple model of
lower complexity. Pixels in complicated regions such as edges and
textures are hard samples. Their residual values are larger, and a more
complicated model is required. To exploit this property, we develop a
simple mechanism to partition pixels based on the variance of their
neighborhood. A pixel whose neighborhood has a smaller variance is an
easy one. Otherwise, it is a hard one. We focus on hard samples for the
rest of this section. The same idea applies to easy samples but the
processing can be greatly simplified. 

\begin{figure}[t]
\centering
\centerline{\includegraphics[width=0.7\linewidth]{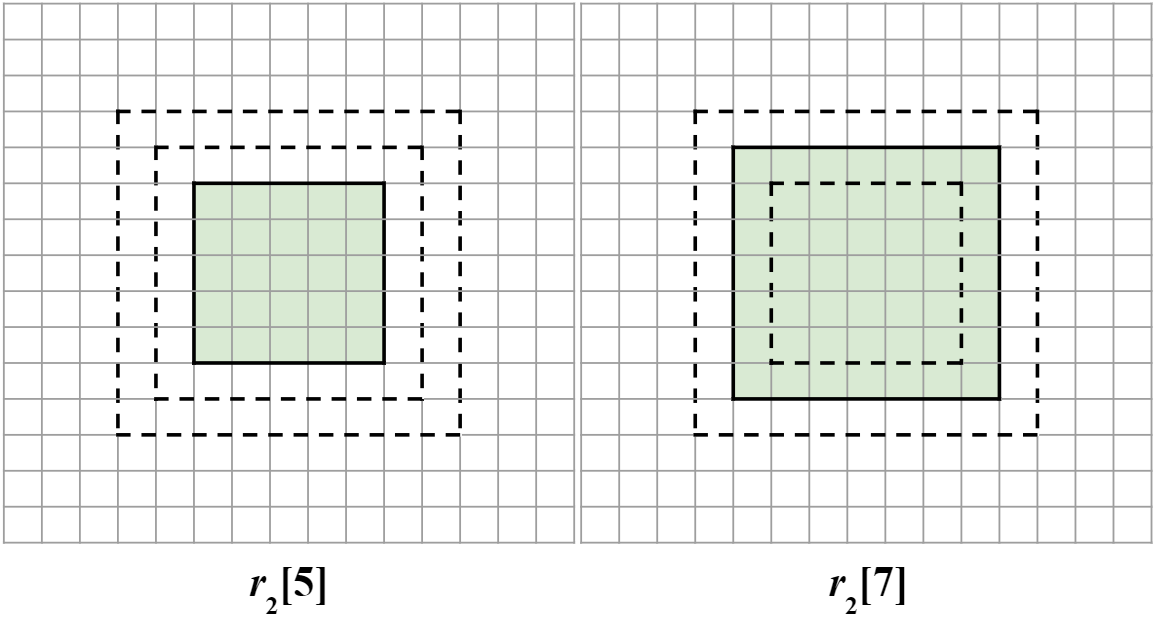}}
\caption{Illustration of central Saab representations of Type 2, where $r_2
[n]$ denotes Representation Type 2 with size $n \times n$.}
\label{fig:type_2}
\end{figure}

\begin{figure}[ht]
\centering
\centerline{\includegraphics[width=1\linewidth]{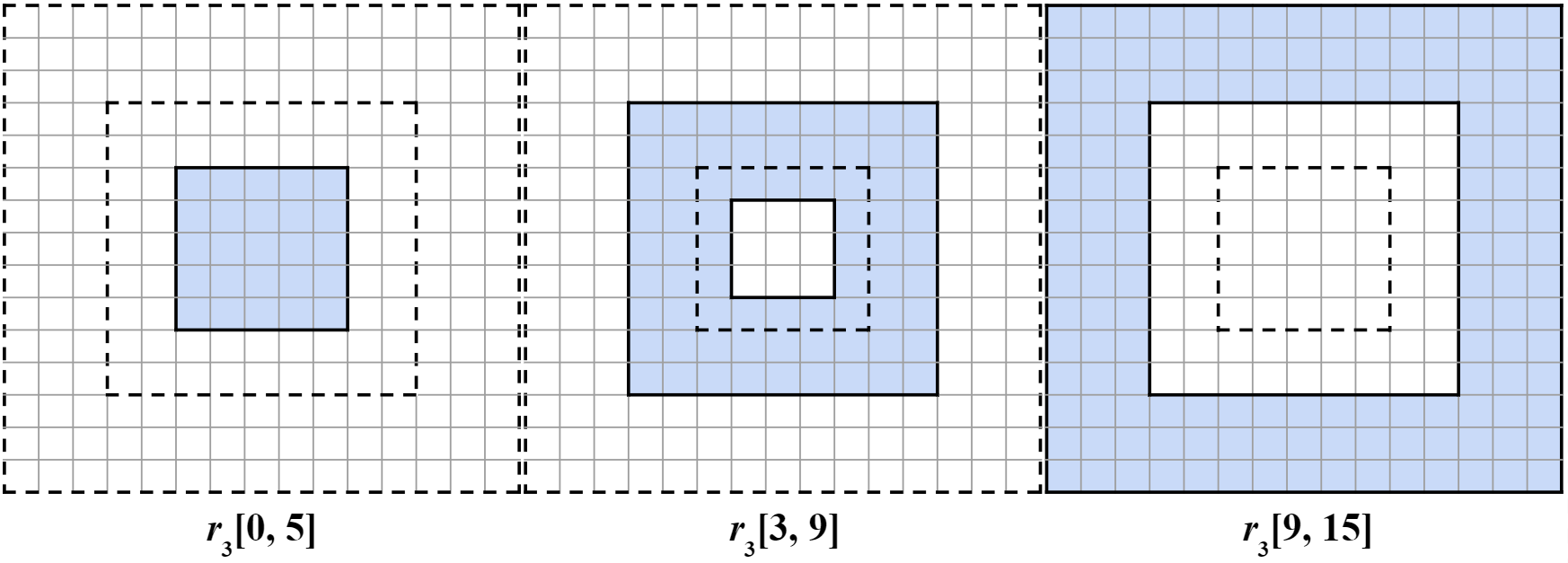}}
\caption{Illustration of ring-wise Saab representations of Type 3,
where $r_3 [n_1,n_2]$, $n_2>n_1$, denotes the difference between the
square of width $n_2$ and the square of width $n_1$.}
\label{fig:type_3}
\vspace{-10pt}
\end{figure}

\subsection{Module 1: Unsupervised Representation Learning} 

The objective of the first module is to generate a rich and diversified
set of representations. Five types of representations are collected with
some justifications. 
\begin{itemize}
\item {\em Type 1: spatial representations.} For a patch of size
$15\times15$, it has 225 pixels values as representations of Type 1. 

\item {\em Type 2: central Saab representations.} Since pixels closer to
the central pixel are more important than distant pixels, we consider
two windows of sizes $5\times5$, $7\times7$ as shown 
in Fig.\ref{fig:type_2} and apply the Saab transform to pixels within each
window to yield the central Saab representations. A window of size $n
\times n$ will yield $n^2$ Saab coefficients as the representations of
Type 2 \footnote{A Saab transform of size $n \times n$ has one fixed
kernel of equal weight $n^{-1/2}$ and $(n^2-1)$ AC kernels that are
derived by the principal component analysis (PCA). We refer to
\cite{kuo2019interpretable} for more details about the Saab transform.}.

\item {\em Type 3: ring-wise Saab representations.} Ring-shaped
neighborhoods are introduced to complement the center-shaped neighborhoods as 
shown in Fig. \ref{fig:type_3}. We apply one-stage Saab transforms with 
$3\times3$ blocks at stride 1 on $r_3 [0, 5]$ and at stride 3 on outter ring 
regions, leading to one DC coefficient and 8 AC coefficients per block. 
These 9-channel responses are decoupled due to PCA.

\item {\em Type 4: Haar filtering followed by channel-wise PCA
representations.} For a neighborhood of size $2\times2$, the Haar
filterbank yields four responses and each response is treated as one
channel. Then, PCA is applied to each channel. Type 4 representation 
is derived from the original Haar response and PCA coefficients. 

\item {\em Type 5: Laws filtering followed by channel-wise PCA
representations.} For a neighborhood of $3\times3$, Laws' filterbank
\cite{laws1980rapid} yields nine responses and each response is treated
as one channel. Then, PCA is applied to each channel. Type 5 representation 
is derived from the original Laws response and PCA coefficients. 

\end{itemize}

\subsection{Module 2: Supervised Feature Learning} 

The total number of representations obtained from Module 1 is around
1500. A subset of the most relevant representations can be selected
based on training data to feed into the regressor. This work adopts a
mechanism called the relevant feature test (RFT)
\cite{yang2022supervised} to achieve this objective. Since RFT is
relatively new, it is briefly reviewed below. Let $[f^i_{min},
f^i_{max}]$ denote the value range of the $i^{th}$ representation, and
partition the samples by a certain value $t$ ($f^i_{min} \le t \le
f^i_{max}$) in the $i^{th}$ representation into two non-overlapping
subsets, denoted by $S^i_L$ and $S^i_R$.  Let $y^i_L$ and $y^i_R$ be the
mean of target values in $S^{i}_L$ and $S^{i}_R$. They are used as the
estimated regression values of all samples in $S^{i}_L$ and $S^{i}_R$,
respectively. The RFT loss is defined as the sum of the estimated
regression MSEs of $S^{i}_L$ and $S^{i}_R$. Mathematically, it is in the
form of
\begin{equation}\label{eq:weighted_MSE}
R^{i}_t = \frac{N^{i}_{L,t} R^{i}_{L,t} + N^{i}_{R,t} R^{i}_{R,t}}{N},
\end{equation}
where $N^{i}_{L,t}$, $N^{i}_{R,t}$, $R^{i}_{L,t}$, and $R^{i}_{R,t}$
are the sample numbers and the estimated regression MSEs in subsets
$S^{i}_L$ and $S^{i}_R$, respectively, and $N=N^{i}_{L,t}+N^{i}_{R,t}$.
The RFT loss function of the $i^{th}$ representation is defined as the
optimized estimated regression MSE over the set, $T$, of all candidate
partition points, i.e.,
\begin{equation}\label{eq:optimized_MSE}
R^i_{op} = \min_{t \epsilon T} R^i_{t}.
\end{equation}
The lower the RFT loss, the better the representation. We compute the
RFT loss values of all representations and sort them in ascending order
to yield an RFT loss curve. The elbow point is considered to select a
subset of representations with lower RFT loss. This set defines the
relevant features to be fed into a regressor in Module 3. 

\subsection{Module 3: Supervised Decision Learning}

In the training process, data augmentation is performed (via 90-degree
rotations and flipping) to enlarge the training sample size, and the
following two options, "clustering" and "prediction" fusion, are
considered. 
\noindent
{\em Clustering.} Perform K-means clustering on ILR patches based
on their HOG features and then train the XGBoost regressor in each
cluster using features selected from Module 2. 
\noindent
{\em Prediction Fusion.} Augment each ILR patch for multiple times,
perform the regression of each one, and take the average of all
prediction results as the ultimate predicted residual value. 

\section{EXPERIMENTS} \label{sec:exp}

\subsection{Experimental Setup} 

The SR experiments are conducted with a scaling factor of 2 and the
BSD200 dataset \cite{martin2001database} is adopted to train the
proposed LSR model.  The tests are performed on four datasets: Set5
\cite{bevilacqua2012low}, Set14\cite{zeyde2012single}, BSD100
\cite{martin2001database}, and Urban100 \cite{huang2015single}.
Following the standard routine, we only process the $Y$ channel for
super resolution. Our model uses the $15\times15$ ILR patches to predict
the residual value of the center pixel of the patch. An ILR patch of
$16\times16$ is also obtained from the $15\times15$ ILR patch by the
Lanczos interpolation for HOG feature extraction. Table
\ref{tab:portion_stat_and_setting} shows statistics and the parameter
setting for easy and hard data samples with notations $RT_h$ and $FU_h$
to represent various representation types and fusion schemes for hard samples. 

\begin{table}[tb]
\vspace{-5pt}
\caption{Statistics and parameter settings for easy and hard
data.}\label{tab:portion_stat_and_setting}
\centering
\footnotesize
\setlength{\tabcolsep}{2.5mm}{
\begin{tabular}{ccc}
\hline
Data Type                     & Easy            & Hard                \\ \hline
Ratio                         & 56\%            & 44\%                \\
Patch variance Range          & $\leq$180       & $\geq$180              \\
Pixel initial MSE(ILR, HR)    & 21.78           & 158.05               \\ \hline
Representation Types                 & [1, 3]               & $RT_h$               \\
Cluster Number                & 1               & 8                   \\ 
XGBoost Regressor Tree Number & 50             & 500                 \\
XGBoost Regressor Max Depth   & 6               & 6                   \\
Fusion                        & No              & Yes ($FU_h$)         \\ \hline
\end{tabular}}
\vspace{-5pt}
\end{table}

\begin{table}[tb]
\vspace{-5pt}
\caption{Average PSNR/SSIM with different settings of representation 
types (RT) for hard data tested on Set5 and Set14.}\label{tab:feature_type}
\centering
\footnotesize
\setlength{\tabcolsep}{2mm}{
\begin{tabular}{cccc}
\hline
PSNR / SSIM & $RT_h$=[1] & $RT_h$=[2] & $RT_h$=[3]         \\ \hline
Set5        & 35.24 / 0.9523 & 36.12 / 0.9576 & 36.16 / 0.9578         \\
Set14       & 31.32 / 0.9074 & 31.94 / 0.9133 & 31.96 / 0.9134         \\ \hline
PSNR / SSIM & $RT_h$=[4] & $RT_h$=[5] & $RT_h$=[1,2,3,4,5] \\ \hline
Set5        & 36.11 / 0.9573 & 36.26 / 0.9583 & 36.34 / 0.9586         \\
Set14       & 31.91 / 0.9130 & 32.04 / 0.9141 & 32.11 / 0.9147         \\ \hline
\end{tabular}}
\vspace{-5pt}
\end{table}

\begin{table}[tb]
\vspace{-5pt}
\caption{Average PSNR/SSIM with different fusion schemes for 
hard data against Set5 and Set14.}\label{tab:fusion_type}
\centering
\footnotesize
\setlength{\tabcolsep}{0.6mm}{
\begin{tabular}{ccccc}
\hline
PSNR/SSIM & $FU_h$=1  & $FU_h$=2  & $FU_h$=3 & $FU_h$=4 \\ \hline
Set5  & 36.33 / 0.9588 & 36.49 / 0.9592 & 36.57 / 0.9595 & 36.60 / 0.9597 \\ 
Set14 & 32.12 / 0.9150 & 32.25 / 0.9156 & 32.30 / 0.9159 & 32.32 / 0.9161 \\ \hline
\end{tabular}}
\vspace{-5pt}
\end{table}

Since the initial MSE of easy data is small, we adopt a simple procedure
to predict the residual values of easy data. For hard data, several
different settings are compared. First, the PSNR/SSIM performance of
different representation types and the union of all five types are
demonstrated in Table \ref{tab:feature_type}. Representation types 2-5
outperform representation types 1 by a clear margin, while the union
of all five types gives the best performance. Here the RFT is applied to
the union of all five representation types below to maintain high
performance. However, one can choose a single representation type such
as type 5 alone to lower the computational complexity. 

Next, we compare the performance of four different decision schemes: 1)
$FU_h=1$: without clustering and prediction fusion, 2) $FU_h=2$: with
clustering but no prediction fusion, 3) $FU_h=3$: with both clustering
and prediction fusion (fusion by 2), and 4) $FU_h=4$: with both
clustering and prediction fusion (fusion by 4). The results are shown in
Table \ref{tab:fusion_type}, which confirm the effectiveness of
clustering and prediction fusion. 

\vspace{-5pt}
\subsection{Quality Performance Comparison} 

\begin{figure*}[t]
\centering
\begin{minipage}{1\linewidth} 
\centerline{\includegraphics[width=1.0\linewidth]{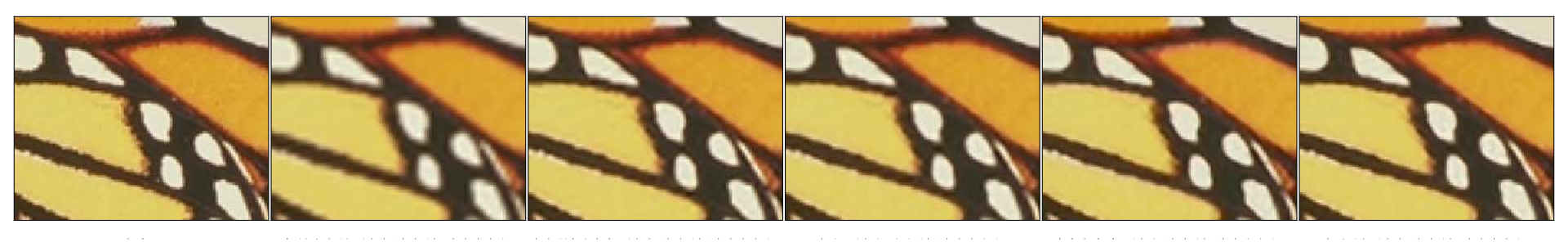}}
\vspace{-5pt}
\end{minipage}
\begin{minipage}{1\linewidth} 
\centerline{\includegraphics[width=1.0\linewidth]{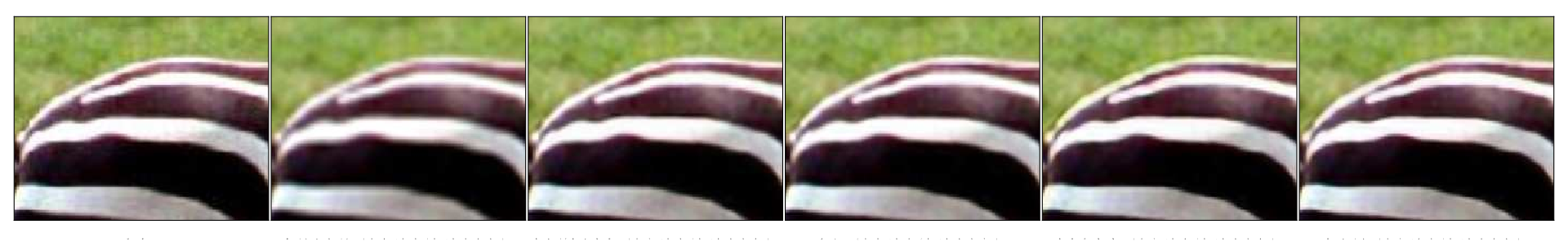}}
\vspace{-5pt}
\end{minipage}
\begin{minipage}{1\linewidth}
\centerline{\includegraphics[width=1.0\linewidth]{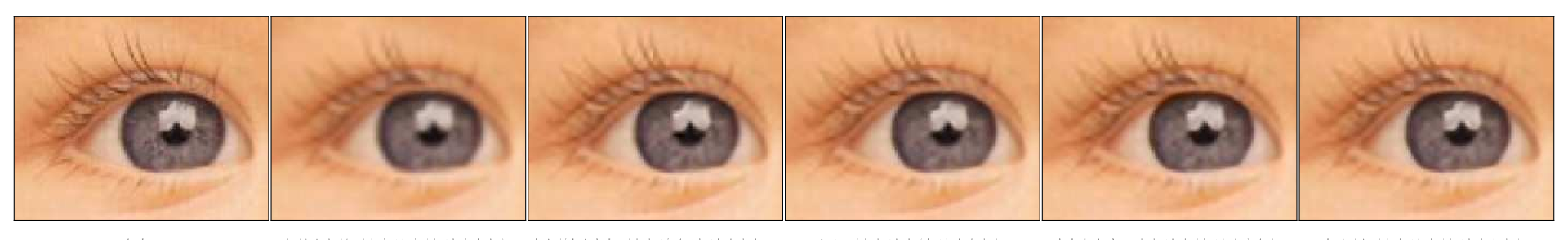}}
\vspace{-5pt}
\end{minipage}
\begin{minipage}{1\linewidth} 
\centerline{\includegraphics[width=1.0\linewidth]{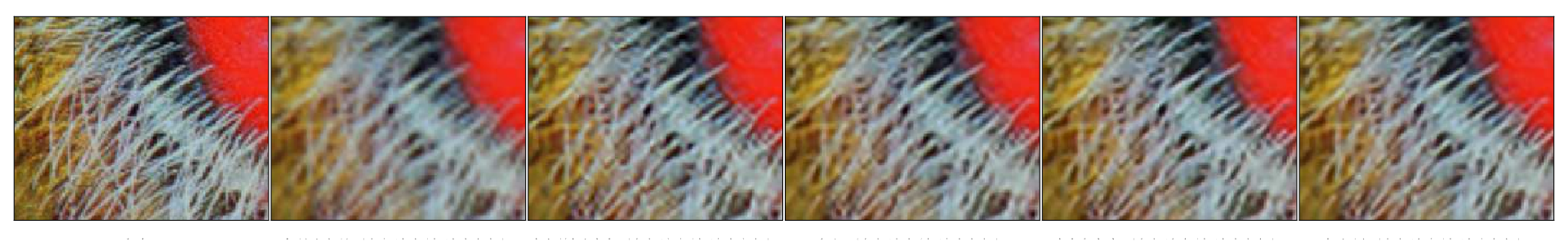}}
\vspace{-7pt}
\end{minipage}
\caption{Visual comparison of 4 test SR images, where the PSNR/SSIM results 
are provided in brackets.}\label{fig:visual_comp}
\vspace{5pt}

\begin{minipage}{1\linewidth} 
\tabcaption{Comparison of averaged PSNR/SSIM values on datasets Set5,
Set14, B100 and Urban100, where red and blue colors indicate the best
and the second best performance for each dataset.}\label{tab:performance_comparison}
\centering
\footnotesize
\setlength{\tabcolsep}{1.5mm}{
\begin{tabular}{cccccc}
\hline
PSNR/SSIM & SelfExSR\cite{huang2015single}       & A+\cite{timofte2014a+}             & SRCNN \cite{dong2015image}    & LSR (Ours), V1    & LSR (Ours), V2  \\ \hline
Set5      & 36.49 / 0.9537 & 36.54 / 0.9544 & {\color{red}{\textbf{36.66}}} / 0.9542 & {\color{blue}{\textbf{36.57}}} / {\color{red}{\textbf{0.9595}}}  & 36.50 / {\color{blue}{\textbf{0.9593}}} \\
Set14     & 32.22 / 0.9034 & 32.28 / 0.9056 & {\color{red}{\textbf{32.42}}} / 0.9063 & {\color{blue}{\textbf{32.30}}} / {\color{red}{\textbf{0.9159}}} & 32.23 / {\color{blue}{\textbf{0.9155}}} \\
BSD100    & 31.18 / 0.8855 & 31.21 / 0.8861 & {\color{blue}{\textbf{31.36}}} / 0.8879 & {\color{red}{\textbf{31.37}}} / {\color{red}{\textbf{0.8985}}} & 31.32 / {\color{blue}{\textbf{0.8980}}} \\
Urban100  & {\color{red}{\textbf{29.54}}} / 0.8967 & 29.20 / 0.8938 & 29.50 / 0.8946 &   {\color{blue}{\textbf{29.51}}} / {\color{red}{\textbf{0.9024}}}   & 29.42 / {\color{blue}{\textbf{0.9013}}}  \\ \hline
\vspace{-15pt}
\end{tabular}}
\end{minipage}

\end{figure*}

Table \ref{tab:performance_comparison} demonstrates the quality
comparison of two versions of the proposed LSR method (V1:
$RT_h$=[1,2,3,4,5], $FU_h$=3, and V2: $RT_h$=[5], $FU_h$=3) and three
light-weight SR methods (SelfExSR\cite{huang2015single}, A+
\cite{timofte2014a+}, and SRCNN\cite{dong2015image}). Note that SRCNN is
the simplest DL-based method. Here we do not include advanced DL-based
solutions in the table since their model sizes and computational
complexity are too high to be used on mobile/edge platforms. This table
exhibits that LSR V1 and LSR V2 achieve the best SSIM performance among
all benchmarking methods while their PSNR performance is close to SRCNN. 

We also show four test SR images in Fig. \ref{fig:visual_comp} for
visual comparison. LSR V1 and SRCNN offer better visual quality with
shaper edges and textures than SelfExSR and A+. Although the visual
quality of LSR and SRCNN is comparable, their complexity is quite
different as presented in the following subsection. 

\begin{table}[tb]
\vspace{-5pt}
\caption{Comparison of computational complexity (FLOPs per pixel) and 
model sizes of five SR methods.}\label{tab:complexity_comparison}
\centering
\footnotesize
\setlength{\tabcolsep}{2mm}{
\begin{tabular}{lrr}
\hline
Complexity     & FLOPs / pixel & Model Size    \\ \hline
A+\cite{timofte2014a+}   & 15.7K(4X)     & 1.06M (18.6X) \\
SRCNN\cite{dong2015image}  & 114K (30X)    & 57.3K (1X)    \\
VDSR\cite{kim2016accurate}           & 1.33M (347X)  & 665K (11.6X)  \\
LSR (Ours), V1 & 9.28K (2.42X) & 774K (13.51X)  \\
LSR (Ours), V2 & 3.83K (1X)    & 770K (13.45X)  \\ \hline
\end{tabular}}
\vspace{-5pt}
\end{table}

\vspace{-5pt}
\subsection{Complexity and Model Size Comparison} 

The computational complexity is measured in terms of floating-point
operations (FLOPs) per pixel in inference, and the model size in terms of the number
of model parameters. Three benchmarking methods are compared to the
proposed LSR in Table \ref{tab:complexity_comparison}.  As one of the
best non-DL-based method, A+ (1024-atom dictionary version) shows
reasonable FLOPs value but large model size. SRCNN (9-5-5 version) has
very small model size, while its FLOPs is large. As a median size
DL-based method, VDSR shows very large FLOPs value.
Although model size of LSR V1 is comparable with VDSR, its FLOPs per pixel is 
only 0.70\% of VDSR. Besides, when achieving similar PSNR/SSIM, The FLOPs per 
pixel of LSR V1 is only 8.11\% of SRCNN. LSR V2 even reduces FLOPs per pixel 
to 3.35\% of SRCNN. Our model LSR shows extremely low inference computaional 
complexity, and its model size is also acceptable for mobile devices.

\vspace{-4pt}
\section{Conclusion and Future Work}\label{sec:conc}

A light-weight SR method, called LSR, was proposed in this work. It
offers good visual quality that is comparable with that of SRCNN, which
is an entry-level DL-based method, but at a significantly lower
computational complexity (i.e., 8.11\% in terms of FLOPs per pixel by V1, even 3.35\% by V2).
Besides, we presented a wide range of design choices that can lower the
computational cost even more with slight quality degradation (see Table
\ref{tab:feature_type} and Table \ref{tab:performance_comparison}).  

There are several topics worth our future research. First, we would like
to generalize our current method to other scale factors ($\times3$, $\times4$, etc.).
Second, it is desired to boost visual quality furthermore while keeping
low computational complexity, which can be achieved by more effective
ensemble learning. Third, it is critical to develop a real-time SR video
solution. The main challenge from image-based to video-based SR is 
preservation of temporal smoothness. 

\section{APPENDICES}\label{appendix}

The detailed calculations on the computational complexity in inference
(in terms of FLOPs) and the model size (in terms of the number of model
parameters) of the methods involved in Table 5 are presented in this
section. The image ``woman.png" of resolution $344 \times 228$ in the
Set5 test dataset is used as an example.  All the calculation is based
on the original codes published by authors of each paper. ``FLOPs per
pixel" in each step is obtained by dividing FLOPs of the whole image by
the pixel number in the final predicted HR.  FLOPs, FLOPs per pixel, and
model sizes are denoted by $F$, $F_p$ and $M$, respectively. 

Interpolation from low resolution images to the same size of high
resolution images is commonly adopted as a pre-processing step in all
algorithms of consideration.  Since the interpolation process is usually
fast and no learned model required, our complexity computation below
does not involve this procedure. Besides, different
algorithms have different strategies in handling image borders. For fair
comparison, we assume that all algorithms generate feature maps and
predict HR image based on the ILR image. 

\subsection{FLOPs and Model Size of Typical Operations}

There are several typical procedures involved in various SR algorithms.
The calculation of $F$ and $M$ on these procedures is discussed below. 

\noindent
{\em Pixel-wise operation.} For a single pixel-wise operation
(addition or multiplication) on a set of number $N$ images (or patches)
with height $H$, width $W$, and depth (number of channels) $C$, we have
$F = H \times W \times C \times N$. 

\noindent
{\em Matrix Multiplication.} For a matrix multiplication between
a $T_h \times T_w$ transform matrix and $T_w \times N$ sample matrix for
$N$ samples, we have
\begin{eqnarray}\label{eq:flops_2d_matrix_multi}
F & = & (2 \times T_w - 1) \times T_h \times N , \\
M & = & T_h \times T_w , \label{eq:modelsize_2d_matrix_multi}
\end{eqnarray}
with $T_w$ multiplications and ($T_w-1$) additions for each element in
the $T_h \times N$ output matrix. 

\noindent
{\em 3D Convolution or Filtering.} For the convolution operations
that generate 2D feature maps by 3D convolution kernels, we use $C_i$ to
denote the number of channels of the input image, and $K_h$ and $K_w$ to
represent height and width of the convolution kernel, respectively. To
generate one spatial feature response in one feature map, we need 
$$
2 \times C_i \times K_h \times K_w - 1
$$ 
operations, including $C_i \times K_h \times K_w$ multiplications and
$C_i \times K_h \times K_w - 1$ additions. The bias-adding operation 
demands one addition operation for each spatial point at a feature map. 
Thus, $F$ and $M$ for a 3D convolution operation without bias on an 
image can be computed as
\begin{eqnarray}
F & = & (2 \times C_i \times K_h \times K_w - 1) \times H_o \times W_o \times C_o,
\label{eq:flops_conv_wo_bias} \\
M & = & (C_i \times K_h \times K_w) \times C_o, \label{eq:modelsize_conv_wo_bias}
\end{eqnarray}
where $H_o$ and $W_o$ are the height and width of the feature maps,
respectively, and $C_o$ is the number of feature maps. If there exists a bias term,
we have
\begin{eqnarray}\label{eq:flops_conv_w_bias}
F & = & (2 \times C_i \times K_h \times K_w) \times H_o \times W_o \times C_o, \\
\label{eq:modelsize_conv_w_bias}
M & = & (C_i \times K_h \times K_w + 1) \times C_o.
\end{eqnarray}

\subsection{A+}

\begin{table}[tb]
\caption{Calculation of FLOPs ($F$), FLOPs per pixel ($F_p$) and model size 
($M$) for A+.}\label{tab:aplus}
\centering
\footnotesize
\vspace{10pt}
\setlength{\tabcolsep}{1.45mm}{
\begin{tabular}{lccccccccc}
\multicolumn{10}{c}{\textbf{ILR Feature Extraction (IFE)}}           \\ \hline
Filter Type  & $C_i$ & $K_h$ & $K_w$ & $H_o$ & $W_o$ & $C_o$ & $F$  & $F_p$ & $M$ \\ \hline
$D^{1}_w$    & 1     & 1     & 3     & 344   & 228   & 1     & 0.39M & 5.00    & 3   \\
$D^{1}_h$    & 1     & 3     & 1     & 344   & 228   & 1     & 0.39M & 5.00    & 3   \\
$D^{2}_w$    & 1     & 1     & 5     & 344   & 228   & 1     & 0.71M & 9.00    & 5   \\
$D^{2}_h$    & 1     & 5     & 1     & 344   & 228   & 1     & 0.71M & 9.00    & 5   \\ \hline
\multicolumn{7}{l}{IFE Sub-total}      & 2.20M & 28.00  &16 \\ \hline
\end{tabular}}

\vspace{10pt}
\setlength{\tabcolsep}{0.95mm}{
\begin{tabular}{lcccccc}
\multicolumn{7}{c}{\textbf{Residue Patch Prediction (RPP)}}       \\ \hline
Step     & $T_h$ & $T_w$ & $N$  & $F$  & $F_p$ & $M$ \\ \hline
ILR Feat. Dim. Red. & 28   & 144    & 18480 & 0.15B  & 1893.43 & 4032 \\
Dist. to ILR Atoms & 1024   & 28 & 18480 & 1.04B & 13270.01   & 28672  \\
Regression Prediction  & 36   & 28   & 18480 & 0.04B   & 466.52 & 1032192 \\ \hline
\multicolumn{4}{l}{RPP Sub-total}      & 1.23B & 15629.96  & 1064896 \\ \hline
\end{tabular}}

\vspace{10pt}
\setlength{\tabcolsep}{1.35mm}{
\begin{tabular}{lccccccc}
\multicolumn{8}{c}{\textbf{HR Image Prediction (HIP)}}              \\ \hline
Step     & H  & W  & C & N  & $F$  & $F_p$ & $M$ \\ \hline
Add. ILR to pred. Res. & 6   & 6   & 1 & 18480 & 0.67M & 8.48  & 0 \\
Cumu. of pixel values  & 6   & 6   & 1 & 18480 & 0.67M & 8.48  & 0 \\
Div. by pixel counter  & 344 & 228 & 1 & 1     & 0.08M  & 1.00  & 0 \\ \hline
\multicolumn{5}{l}{HIP Sub-total}      & 1.41M & 17.96  & 0 \\ \hline
\end{tabular}}

\vspace{10pt}
\setlength{\tabcolsep}{2mm}{
\begin{tabular}{lccc}
\multicolumn{4}{c}{\textbf{Summary}}              \\ \hline
Procedure       & $F$      & $F_p$         & $M$  \\ \hline
IFE Sub-total   & 2.20M    & 28.00         & 16         \\
RPP Sub-total   & 1.23B    & 15629.96      & 1064896    \\
HIP Sub-total   & 1.41M    & 17.96         & 0          \\ \hline
Total           & 1.23B    & 15675.93      & 1064912    \\ \hline
\end{tabular}}
\end{table}

As an example-based SISR algorithm, A+ uses $6 \times 6$ ILR patches
(with overlapping width 2) to predict the the corresponding $6 \times 6$
residue patches, and generate the average values for patch overlapping
regions. A+ mainly contains three sequential procedures: 1) ILR feature
extraction, 2) residue patch prediction, and 3) HR image prediction.
The calculation of $F$, $F_p$ and $M$ of A+ are given in Table
\ref{tab:aplus}. Based on the example image, 18480 $6 \times 6$ patches
are formed in the entire process. 

{\bf ILR Feature Extraction.} Four feature maps are generated using the
first and the second order derivative filters ($D^1$, $D^2$) along the
image height ($D^1_h$, $D^2_h$) and width ($D^1_w$, $D^2_w$),
respectively. 

{\bf Residue Patch Prediction.} For a $6 \times 6$ ILR patch, ILR raw
features of 144 dimensions are formed by taking the corresponding $6
\times 6$ region in four feature maps and conduct feature concatenation.
Afterward, three steps are executed to generate the residue patches: 1)
reduce ILR features from 144 dimensions to 28 dimensions, 2) calculate
distance to 1024 ILR dictionary atoms to identify the closest atom for
each ILR patch, and 3) predict the residue patch values by the regressor
associated with the closest ILR dictionary atom for each ILR patch. All
the three steps are implemented by 2D matrix multiplication.  Then, raw
regression predictions (36-D) are reshaped into $6 \times 6$ patches to
form the eventual residue patch prediction. In the third step,
"Regression Prediction", of this procedure, $F$ is calculated by the
closest regressor, while $M$ includes regressors associated all 1024 ILR
dictionary atoms. 

{\bf HR Image Prediction.} Predicted HR patches are obtained by adding
corresponding ILR values to the predicted residue patches. Then, they
are used to reconstruct the complete predicted HR image one by one. Two
$344 \times 228$ all-zero matrices are generated, with one for pixel
value accumulation, and the other for counting pixel coverage times from
different patches. The final HR image prediction is obtained by the
division of the accumulation matrix by the counting matrix. All steps in
this procedure are pixel-wise operations. 

\subsection{SRCNN}

SRCNN is a DL-based method that has three convolution layers with bias
on ILR images to generate predicted HR images. The calculation of $F$,
$F_p$ and $M$ are shown in Table \ref{tab:srcnn}. 

\begin{table}[tb]
\caption{Calculation of FLOPs ($F$), FLOPs per pixel ($F_p$) and model
size ($M$) for SRCNN.}\label{tab:srcnn}
\centering
\footnotesize
\vspace{3pt}
\setlength{\tabcolsep}{1.5mm}{
\begin{tabular}{lccccccccc}
\hline
Steps & $C_i$ & $K_h$ & $K_w$ & $H_o$ & $W_o$ & $C_o$ & $F$  & $F_p$ & $M$ \\ \hline
conv1 & 1    & 9    & 9    & 344  & 228  & 64   & 0.81B  & 10368  & 5248  \\
conv2 & 64   & 5    & 5    & 344  & 228  & 32   & 8.03B & 102400 & 51232  \\
conv3 & 32   & 5    & 5    & 344  & 228  & 1    & 0.13B  & 1600   & 801 \\ \hline
Total &  &  &  &  &  &  & 8.97B & 114368        & 57281      \\ \hline
\end{tabular}}
\end{table}

\subsection{VDSR}

VSDR is a 20-layer DL-based method with $3 \times 3$ kernel for each
layer. VDSR utilizes ILR images to predict residue images, and final
predicted HR images are obtained by adding predicted ILR residues to the
ILR images. The calculation explanatioin of $F$, $F_p$ and $M$ are
exhibited in Table \ref{tab:vsdr}, with $N_l$ denotes number of layers. 

\begin{table}[tb]
\caption{Calculation of FLOPs ($F$), FLOPs per pixel ($F_p$) and model
size ($M$) for VDSR.}\label{tab:vsdr}
\centering
\footnotesize
\vspace{3pt}
\setlength{\tabcolsep}{0.65mm}{
\begin{tabular}{lcccccccccc}
\hline
Steps          & $C_i$ & $K_h$ & $K_w$ & $H_o$ & $W_o$ & $C_o$ & $N_l$ & $F$  & $F_p$ & $M$ \\ \hline
conv1          & 1    & 3    & 3    & 344  & 228  & 64   & 1   & 90.35M     & 1152          & 576        \\
conv2 - 19 & 64   & 3    & 3    & 344  & 228  & 64   & 18   & 104.09B & 1327104       & 663552     \\
conv20         & 64   & 3    & 3    & 344  & 228  & 1    & 1  & 90.35M     & 1152          & 576        \\
post-process   &      &      &      & 344  & 228  & 1    & 1   & 78.43k        & 1             & 0          \\ \hline
Total          &      &      &      &      &      &      &    & 104.29B & 1329409       & 664704     \\ \hline
\end{tabular}}
\end{table}

\begin{table}[bth]
\caption{Calculation of FLOPs per pixel ($F_p$) and model size ($M$) for LSR
in each module.}\label{tab:ours_modulewise}
\centering
\footnotesize
\vspace{10pt}
\setlength{\tabcolsep}{1.5mm}{
\begin{tabular}{lccccccc}
\multicolumn{8}{c}{\textbf{Module 1: Unsupervised Representation Learning (URL)}}                                    \\ \hline
$RT$  & $C_i$ & $K_h$ & $K_w$ & $C_o$ & $N_{type}$ & $F_p$ & $M$ \\ \hline
Type 1, Spatial    & 1    & 1    & 1  & 1    & 0   & 0     & 0    \\
Type 2, Central Saab & 1  & 5  & 5  & 25   & 1   & 1225  & 625  \\
Type 2, Central Saab & 1  & 7  & 7  & 49   & 1   & 4753  & 2401 \\
Type 3, Ringwise Saab & 1  & 3  & 3  & 9   & 1   & 153  & 81 \\
Type 4, Haar \& PCA   & 1  & 2  & 2  & 4   & 2   & 56   & 32  \\
Type 5, Laws \& PCA   & 1  & 3  & 3  & 9   & 2   & 306  & 162 \\ \hline
\end{tabular}}

\vspace{3pt}
\setlength{\tabcolsep}{1mm}{
\begin{tabular}{lccccc}
\hline
URL Sub-total  & $RT$ & $F_p / f$ & $f$ & $F_p$ & $M$ \\ \hline
Easy           & [1, 3]    & 153   & 1 & 153 & 81    \\
Hard (V1)      & [1, 2, 3, 4, 5]  & 6493  & 2 & 12986 & 3301  \\ 
Hard (V2)      & [5]  & 306  & 2  & 612 & 102  \\ \hline
\end{tabular}}

\vspace{10pt}
\setlength{\tabcolsep}{2mm}{
\begin{tabular}{lcccc}
\multicolumn{5}{c}{\textbf{Module 2: Supervised Feature Learning (SFL)}} \\ \hline
RFT Sub-total      & $RT$ & $N_{fr}$ & $F_p$ & $M$ \\ \hline
Easy               & [1, 3]       & 105    & 0    & 105    \\
Hard (V1)          & [1, 2, 3, 4, 5] & 374 & 0    & 374    \\ 
Hard (V2)          & [5]          & 135    & 0    & 135    \\ \hline
\end{tabular}}

\vspace{10pt}
\setlength{\tabcolsep}{2mm}{
\begin{tabular}{lcccc}
\multicolumn{5}{c}{\textbf{Module 3: Supervised Decision Learning (SDL)}}    \\
\vspace{-10pt}
\end{tabular}}

\setlength{\tabcolsep}{2.9mm}{
\begin{tabular}{lcccccc}
\hline
Cluster Pred. & $N_{fc}$ & $N_c$ & $F_p$ / $f$ & $f$ & $F_p$ & $M$   \\ \hline
Easy      & 0     & 1    & 0    & 1  & 0    & 0   \\
Hard      & 32    & 8    & 760  & 2  & 1520 & 256 \\ \hline
\end{tabular}}
\setlength{\tabcolsep}{0.9mm}{

\vspace{3pt}
\begin{tabular}{lcccccccc}\hline
Regressor Pred. & $N_{tree}$ & $d_M$ & $F_p$/$f$ & $f$ & $F_p$ & $M$/cluster & $N_c$ & $M$      \\ \hline
Easy    & 50   & 6   & 300    & 1   & 300  & 9500   & 1  & 9500   \\
Hard    & 500  & 6   & 3000   & 2   & 6000 & 95000  & 8  & 760000 \\ \hline
\end{tabular}}

\vspace{3pt}
\setlength{\tabcolsep}{4.5mm}{
\begin{tabular}{lccc}
\hline
Prediction Fusion  & $f$ & $F_p$ & $M$ \\ \hline
Easy           & 1 & 0 & 0  \\
Hard           & 2 & 2 & 0  \\ \hline
\end{tabular}}

\vspace{3pt}
\setlength{\tabcolsep}{5.5mm}{
\begin{tabular}{lcc}
\hline
SDL Sub-total  & $F_p$ & $M$ \\ \hline
Easy           & 300  & 9500    \\
Hard           & 7522 &	760256  \\ \hline
\end{tabular}}
\end{table}

\begin{table}[tb]
\caption{Calculation of FLOPs and model size for VDSR.}\label{tab:ours_summary}
\centering
\footnotesize
\vspace{10pt}
\setlength{\tabcolsep}{3mm}{  
\begin{tabular}{ccccc}
\multicolumn{5}{c}{\textbf{V1 Summary}}                                                                  \\ \hline
\multicolumn{1}{c|}{Complexity}    & \multicolumn{2}{c|}{$F_p$} & \multicolumn{2}{c}{$M$} \\ \hline
\multicolumn{1}{c|}{Data Type}     & Easy  & \multicolumn{1}{c|}{Hard}  & Easy          & Hard           \\
\multicolumn{1}{c|}{URL Sub-total} & 153   & \multicolumn{1}{c|}{12986} & 81            & 3301           \\
\multicolumn{1}{c|}{SFL Sub-total} & 0     & \multicolumn{1}{c|}{0}     & 105           & 374            \\
\multicolumn{1}{c|}{SDL Sub-total} & 300   & \multicolumn{1}{c|}{7522}  & 9500          & 760256         \\
\multicolumn{1}{c|}{Post-process}  & 1     & \multicolumn{1}{c|}{1}     & 0             & 0              \\ \hline
\multicolumn{1}{c|}{Sub-Total}     & 454   & \multicolumn{1}{c|}{20509} & 9686          & 763931         \\
\multicolumn{1}{c|}{$w$}        & 0.56  & \multicolumn{1}{c|}{0.44}   & 1             & 1              \\ \hline
\multicolumn{1}{c|}{Total}         & \multicolumn{2}{c|}{9278}          & \multicolumn{2}{c}{773617}     \\ \hline
\end{tabular}}

\vspace{10pt}
\setlength{\tabcolsep}{3mm}
{\begin{tabular}{ccccc}
\multicolumn{5}{c}{\textbf{V2 Summary}}                                                                  \\ \hline
\multicolumn{1}{c|}{Complexity}    & \multicolumn{2}{c|}{$F_p$} & \multicolumn{2}{c}{$M$} \\ \hline
\multicolumn{1}{c|}{Data Type}     & Easy  & \multicolumn{1}{c|}{Hard}  & Easy          & Hard           \\
\multicolumn{1}{c|}{URL Sub-total} & 153   & \multicolumn{1}{c|}{612}   & 81            & 162            \\
\multicolumn{1}{c|}{SFL Sub-total} & 0     & \multicolumn{1}{c|}{0}     & 105           & 135            \\
\multicolumn{1}{c|}{SDL Sub-total} & 300   & \multicolumn{1}{c|}{7522}  & 9500          & 760256         \\
\multicolumn{1}{c|}{Post-process}  & 1     & \multicolumn{1}{c|}{1}     & 0             & 0              \\ \hline
\multicolumn{1}{c|}{Sub-Total}     & 454   & \multicolumn{1}{c|}{8135}  & 9686          & 760553         \\
\multicolumn{1}{c|}{$w$}        & 0.56  & \multicolumn{1}{c|}{0.44}  & 1             & 1              \\ \hline
\multicolumn{1}{c|}{Total}         & \multicolumn{2}{c|}{3834}          & \multicolumn{2}{c}{770239}     \\ \hline
\end{tabular}}
\end{table}

\subsection{LSR}

Since inference samples are partitioned into easy and hard samples, we
compute FLOPs per pixel, $F_p$, using the weighted sum of easy and hard
samples, where the weight is determined by the ratio of easy and hard
samples in representative images. The model size includes the model
parameters in both partitions. $F_p$ and $M$ calculations for each
module are provided in Table \ref{tab:ours_modulewise}. The complexity
calculation for V1 and V2 are, respectively, summarized in Table
\ref{tab:ours_summary}. 

{\bf Module 1: Unsupervised Representation Learning (URL).} Being slightly
different from the 3D convolution operation without bias (eq.
\ref{eq:flops_conv_wo_bias}) using 3D kernels or filters, LSR uses
channel-wise 2D filters. $RT$ in Table \ref{tab:ours_modulewise}
denotes the representation type(s). $F_p$ and $M$ of LSR in Module 1 for
a certain representation type are obtained by
\begin{eqnarray}\label{eq:flops_ours_url}
F_p & = & C_i \times (2 \times K_h \times K_w - 1) \times C_o \times N_{type}, \\
M & = & C_i \times (K_h \times K_w) \times C_o \times N_{type}, \label{eq:modelsize_ours_url}
\end{eqnarray}
where $N_{type}$ counts the type number regarding to original filter or
PCA filters for a certain representation type, which is typically used
for Type 4 and 5 Representation. $N_{type} = 0$ means no filtering
operation needed, nor is filter parameter needed to store. $N_{type} =
1$ means the normal transform operations and corresponding filters are
required. $N_{type} = 2$ means both normal transform and channel-wise
PCA transform involved.  We use $f$ to denote inference augmentation
times for fusion. The sibling candidate samples for one inference sample
undergo the complete prediction process. Thus, $F_p$ for one inference
sample needs to consider the $f$ factor. 

{\bf Module 2: Supervised Feature Learning (SFL).} $F$ in this module is
zero due to absence of mathematical operations. The model stores the
representation indices which are selected as regression features. Thus,
$M$ is determined by the number of features for regression ($N_{fr}$)
selected from representation pool. 

{\bf Module 3: Supervised Decision Learning (SDL).} There are three
steps for inference samples in this module: cluster prediction,
regressor prediction, and prediction fusion. 
\begin{itemize}
\item {\em Cluster Prediction.} Denote the number of clustering feature
by $N_{fc}$, and cluter number by $N_c$. For one sample, the FLOPs
consumed in its cluster label prediction derives from the calculation of
L2 distance to all $N_c$ cluster centroids, which can be simplified as
first term in eq. (\ref{eq:flops_ours_sdl_clst}). 
\begin{equation}\label{eq:flops_ours_sdl_clst}
F_p = max((3 \times N_{fc} - 1) \times N_c, 0) \times f,
\end{equation}
including $N_{fc}$ subtraction, $N_{fc}$ multiplication, and
$(N_{fc}-1)$ addition operations with respect to each cluster centroid.
The maximum operation in eq. (\ref{eq:flops_ours_sdl_clst}) is for the
calculation generalization for easy data without clustering procedure.
$F_p$ for each inference sample involves the multiplicaiton by $f$
number of sibling candidate samples. $M$ only contains all cluster 
centroids in the clustering procedure. It is equal to
\begin{equation}\label{eq:modelsize_ours_sdl_clst}
M = N_{fc} \times N_c.
\end{equation}

\item {\em Regressor Prediction.} LSR learns an XGBoost regressor in each
cluster for prediction. The upper bound of FLOPs of one sample
prediction by a XGBoost regressor with $N_{tree}$ number of boosting
trees and maximum depth $d_M$ is calculated by $d_M \times N_{tree}$,
where $d_M$ is the FLOPs value for one boosting tree, as one sample at
most traverses $d_M$ nodes until it arrives at one leaf node, and one node
only performs one operation. Similar to the clustering procedure, FLOPs
for one inference sample also need the multiplication by fusion number
$f$. Thus, we have
\begin{equation}\label{eq:flops_ours_sdl_reg}
F_p = d_M \times N_{tree} \times f,
\end{equation}

For a complete binary decision tree with depth $d_M$, the numbers of
leaf nodes ($N_{leaf}$) and parent nodes ($N_{parent}$) are calculated by
\begin{equation}\label{eq:modelsize_ours_sdl_reg_node}
N_{leaf} = 2^{d_M},
N_{parent} = \sum_{d=1}^{d_M-1} 2^{d} =  2^{d_M} - 1,
\end{equation}
and its parameter number is $(2 \times N_{parent} + N_{leaf})$, with
parent nodes storing feature index and partition threshold, and leaf
nodes storing the prediction weight. Thus, the value of $M$
for all regressors is bounded by
\begin{equation}\label{eq:modelsize_ours_sdl_reg}
M = (2 \times N_{parent} + N_{leaf}) \times N_{tree} \times N_c.
\end{equation}

\item {\em Prediction Fusion.} Hard data need additional ($f-1$)
additional operation and one division operation to average the
predictions from sibling samples. 

\end{itemize}

The raw regression results are residue pixel values. We need a
pixel-wise post-processing step that adds a residual to a ILR value to
yield the ultimate HR prediction. The difference between V1 and V2
models lies in the URL representation preparation and SFL feature
selection. 

\clearpage
\newpage
\bibliographystyle{IEEEtran}
\bibliography{refs}

\end{document}